\documentclass[aps,prd,twocolumn]{revtex4}
\usepackage[T2A]{fontenc}
\usepackage[cp1251]{inputenc}
\usepackage[english]{babel}
\usepackage{amsmath}
\usepackage{amsbsy}
\usepackage{amssymb}
\usepackage{graphicx}
\usepackage{calrsfs}
\usepackage[hypertex]{hyperref}

\DeclareMathAlphabet\mathbfcal{OMS}{cmsy}{b}{n}

\newcommand{\apjl}{{\em Ap.~J. Letters{\rm}}}

\newcommand{\mnras}{{\em M.N.R.A.S.{\rm}}}

\begin{document}

\def\b{\boldsymbol}
\def\d{\partial}
\def\p{\varpi}
\def\e{\varepsilon}
\def\k{\kappa}
\def\ds{\displaystyle}
\def\t{\tilde}
\def\apjs{ApJS}

\title{Polarization of radiation of electrons in highly turbulent magnetic fields}

\author{A.Yu.~Prosekin}
\email{Anton.Prosekin@mpi-hd.mpg.de}
\affiliation{Max-Planck-Institut f\"ur Kernphysik,
Saupfercheckweg 1, D-69117 Heidelberg, Germany}

\author{S.R.~Kelner}
\email{Stanislav.Kelner@mpi-hd.mpg.de}
\affiliation
{Max-Planck-Institut f\"ur Kernphysik,
Saupfercheckweg 1, D-69117 Heidelberg, Germany}
\affiliation{Research Nuclear University (MEPHI), Kashirskoe shosse 31,
115409 Moscow, Russia}

\author{F.A.~Aharonian}
\email{Felix.Aharonian@mpi-hd.mpg.de}
\affiliation{Dublin Institute for Advanced Studies, 31 Fitzwilliam Place,
Dublin 2, Ireland}
\affiliation
{Max-Planck-Institut f\"ur Kernphysik,
Saupfercheckweg 1, D-69117 Heidelberg, Germany}
\affiliation{Research Nuclear University (MEPHI), Kashirskoe shosse 31,
115409 Moscow, Russia}

\date{\today}

\begin{abstract}
We study the polarization properties of the jitter and synchrotron radiation
produced by electrons in highly turbulent anisotropic magnetic fields.
The net polarization is provided by the geometry of the magnetic field the directions of which are parallel to a certain plane. Such conditions
may appear in the relativistic shocks during the amplification of the magnetic field 
through the so-called Weibel instability. While the polarization properties of the jitter
radiation allows extraction of direct information on the turbulence spectrum as well as
the geometry of magnetic field, the polarization of the synchrotron radiation
reflects the distribution of the magnetic field over its strength. For the isotropic distribution of monoenergetic electrons, we found that the degree
of polarization of the synchrotron radiation is larger than the polarization of the jitter radiation. For the power-law energy distribution of electrons the relation between the degree of polarization of synchrotron and jitter radiation depends on the spectral index of the distribution.  
\end{abstract}

\maketitle
\section{Introduction}
Turbulent magnetic fields play an important role in many
astrophysical processes such as amplification of magnetic fields,
accretion, viscous heating and thermal conduction
in a turbulent magnetised plasma, etc. \citep{Schekochihin2007}. 
One of the most important processes where the presence of turbulent magnetic
fields is necessary is the diffusive shock acceleration.
In particular, the collisionless shock waves themselves are generated via magnetic turbulence, which
mediates interactions between particles. In the acceleration process the turbulence is required to trap accelerated
particles around shock front for successive crossings, in which they gain energy.

The acceleration of the particles is accompanied by their radiation. The character of the 
radiation can reveal details of the acceleration as well as the properties of the turbulent
medium where the acceleration occurs. The geometry and the scale of turbulence could be reflected
in the polarization properties of the radiation. A completely isotropic turbulence does not produce
any net polarization. However, if the scale of the turbulence is sufficiently large one can detect the fluctuations of the polarization and study the structure of the magnetic field \citep{Bykov2009}. In the case of small-scale turbulence, the only way to observe a polarised radiation is the specific anisotropic geometry of the turbulent magnetic field. This concerns, for example, the objects like GRBs.

The turbulent magnetic fields in the shock waves can be generated by a variety of plasma instabilities
\citep{Bret2004,Lemoine2010}. Particle-in-cell (PIC) simulations show that a Weibel instability is
a key component for generation of the relativistic collisionless shock waves and amplification of
magnetic fields \citep{Spitkovsky2008,Martins2009,Medvedev2011}. As proposed in Ref.~\cite{Medvedev1999} 
in the context of the magnetic field amplification in GRBs, the
anisotropy of colliding beams is transferred to the energy of the small-scale turbulent magnetic field. The
characteristic scale of the turbulence is of the order of the plasma skin-depth. Here it is interesting to note that
the amplification occurs predominantly for the components of the field which are perpendicular to the direction
of beams. The analysis of PIC simulations conducted in Ref.~\cite{Medvedev2011}  reveals also a significant anisotropy
of the turbulence at the  saturation stage of amplification. However, the
turbulence becomes more isotropic in a course of non-linear evolution far behind the shock front \cite{Medvedev2011}.

The radiation of electrons in the small-scale magnetic fields significantly differs from the regular
synchrotron radiation \cite{Medvedev2000}. Because of the turbulence, the formation length of the radiation can be smaller than the formation length required for synchrotron radiation. In this case the qualitatively different type of radiation - jitter radiation - appears. The conditions for realization of this regime of radiation is the smallness of the
characteristic length of the turbulence $\lambda$ compared to the {\it non-relativistic} Larmor radius
$R_L=\frac{mc^2}{eB}$. Thus the appearance of the jitter radiation is determined solely by the properties of 
the magnetic field. Because of smaller formation length, the characteristic frequency of jitter radiation $\omega_j$
is larger than the characteristic frequency of synchrotron radiation $\omega_0$ by the factor $\delta_j=\lambda/R_L$, i.e. $\omega_j=\omega_0/\delta_j$,  where $\delta_j\ll 1$. 

The mechanism of jitter radiation has been revisited in Ref.~\cite{Kelner2013}. It has been shown that the power spectrum of radiation behaves as a constant at frequencies smaller than the characteristic frequency of jitter
radiation, contrary to the earlier claimed $\sim\omega^1$ behaviour \cite{Medvedev1999}. The spectrum at 
high frequencies is a power-law with index determined by the turbulent spectrum. The total power of the 
jitter radiation equals the total power of the synchrotron radiation in the isotropic magnetic field. Thus presence
of the small-scale turbulence affects the radiation spectrum but does not touch total losses. In this paper we show
that a similar relation occurs for polarisation properties: the turbulence influences only the spectral degree
of polarisation whereas the degree of polarisation of total radiation is the same for jitter and synchrotron
radiations.

Concerning the polarisation of radiation from GRBs, the synchrotron radiation is assumed to be the main mechanism for
the production of polarised emission. While the Weibel instability seems to be inevitable in
generation of inner and outer shocks, it is not clear whether the Weibel instability indeed
produces small enough turbulence for the operation of the jitter regime. The simulations of Ref.~\cite{Medvedev2011} show that the jitter radiation regime can operate during the growing stage while the current filaments
merge, and, at the later times, after reaching the non-linear regime when magnetic field starts to decay. While the 
growth of the magnetic field occurs very fast, the decaying stage looks more promising for production of significant
portion of jitter radiation. At the saturation stage, it is more probable that the synchrotron regime is at work.
So it makes a sense to consider the polarisation properties  of both synchrotron and jitter regimes for a specific
configuration of the turbulent magnetic field generated by Weibel instability.

The polarisation properties of the radiation in the turbulent magnetic field with the so-called slab geometry have
been studied in refs.~\cite{Laing1980,Mao2013}. Assuming an independence of the radiation from
different parts of the emitting region, Laing \cite{Laing1980} has averaged the radiation from power-law electron
distribution over the isotropically distributed directions of the magnetic field in the plane. While this
approach works for synchrotron radiation, the calculations in the case of jitter radiation should take into account
the coherence of the turbulent magnetic field. This has not been done in the calculations of Ref.~\cite{Mao2013}
where the polarisation of jitter radiation has been studied in the manner of Ref.~\cite{Laing1980}. It has led them to
incorrect the conclusion that jitter radiation can give a $100\%$ polarisation.

In this paper we study the polarisation properties of the radiation from isotropically distributed electrons
in a turbulent magnetic field with slab geometry. The calculations have been conducted in the general tensor form
which is not attached to any specific coordinate system and allows us to avoid any assumptions about principle
axis of the polarisation ellipse. For calculations we follow the approach proposed in Ref.~\cite{Kelner2013}. The
averaging of the obtained formulae for jitter radiation over all directions of observation reproduces the results
obtained for the isotropic turbulence considered in Ref.~\cite{Kelner2013}. The derived formulae can be used for an
arbitrary energy distribution of electrons. We compare the polarisation properties of the synchrotron and jitter
radiation and show that the synchrotron radiation is more polarised in the case of monoenergetic distributions. For
the power-law distribution of electrons the analytical formula for the degree of polarisation of jitter radiation
is obtained.

The paper has the following structure. Sections 2 and 3 describe the calculations of the polarization 
produced in the jitter and synchrotron radiation regimes, respectively. In the Section 4 we present the results and
compare two cases. Finally, in Section 5 we discuss the main results and make conclusions.

\section{Polarization of jitter radiation}
The spectral power of the emission of a charged particle at the moment $t$ can be written
in the form \citep{Kelner2013,Jackson1998}:
\begin{equation}\label{eq:pow}
P_{\b n \omega}(t)=\frac{e^2}{4\pi^2c}\int\limits_{-\infty}^{\infty}\b U(t+\tau/2)\b U^*(t-\tau/2)d\tau,
\end{equation}
where
\begin{equation}
\b U(t)=\frac{\b n \times \left[(\b n-\b \beta(t))\times \dot{\b \beta}(t)\right]}{(1-\b n\b \beta(t))^2}
e^{i\Phi(t)}.
\end{equation}
The complex vector $\b U(t)$ depends on time through the particle velocity $\b \beta(t)=\b v(t)/c$, the
acceleration $\dot{\b\beta}(t)$, and the exponential factor $\Phi(t)=\omega(t-\b n\b r(t)/c)$, which depends on the
radius-vector of the particle. Here $\b n$ is the unit vector in the direction of observation. The function
$\b U^*(t)$ is the complex conjugation of $\b U(t)$.

Eq.~(\ref{eq:pow}) gives the total power of radiation per unit frequency and per unit solid angle irrespective of polarization. To take into account the directions of the oscillation of the electric field vector in the electromagnetic wave, one should consider instead of just the scalar product the direct product of the electric field vector in the Fourier space. This can be done by rewriting Eq.~(\ref{eq:pow}) in the tensor form:
\begin{equation}\label{eq:powik}
(P_{\b n \omega}(t))_{ik}=\frac{e^2}{4\pi^2c}\int\limits_{-\infty}^{\infty}U_{i}(t+\tau/2)
U_{k}^*(t-\tau/2)d\tau.
\end{equation}
The intensity of the radiation polarized in the direction $\b e$ then can be easily found as $(P_{\b n \omega}(t))_{ik}e_ie_k$. Note that the quantity $(P_{\b n \omega}(t))_{ik}$ is the equivalent of the polarization tensor which is usually determined as $(P_{\b n \omega}(t))_{ik}$ normalised to the intensity $P_{\b n \omega}(t)=(P_{\b n \omega}(t))_{ii}$ \citep{Landau2}. In this paper, for convenience, we refer to this quantity as polarization tensor.

The jitter radiation is realized in the small-scale chaotic magnetic field with
the correlation length $\lambda$ smaller that nonrelativistic Larmor radius $R_L=mc^2/eB$. We assume that
statistically averaged magnetic field is $\langle\b B\rangle=0$. The radiation should be averaged over all
possible configurations of magnetic field. Under this condition
$\lambda$ determines the scale of particle path where the observed radiation is generated. 
At distances $c\tau\gg\lambda$, the particle radiation is incoherent. The value of the
integrand in Eq.~(\ref{eq:powik}) tends to zero at $\tau\gg\lambda/c$ because the magnetic fields
become uncorrelated and the time-averaged product $\langle B_{1\rho}B_{2\sigma}\rangle=\langle B_{1\rho}\rangle
\langle B_{2\sigma}\rangle=0$.

As discussed in Ref.~\cite{Kelner2013} , the condition  $\lambda\ll R_L$ allows us to perform the
calculations in the framework of perturbation theory. Since $\dot{\b\beta}(t)=e(\b\beta\times\b
B)/(mc\gamma)$, we can assume, as a zeroth approximation, that $\beta(t\pm \tau/2)=\beta(t)=\beta$, 
$\b r(t+\tau)=\b r(t)\pm \b \beta\tau/2$. With such accuracy, we have

\begin{equation}
(P_{\b n \omega}(t))_{ik}=\frac{e^2}{4\pi^2c}\left(\frac{e}{mc\gamma\eta^2}\right)^2
\int\limits_{-\infty}^{\infty}p^{+}_{i}p^{-}_{k}e^{i\omega\eta\tau}d\tau,
\end{equation}
where $\eta=1-\b n\b \beta$, $\b p^{\pm}=\b n \times \left[(\b n-\b \beta)\times \left(\b\beta\times\b
B^\pm\right)\right]$, $\b B^{\pm}=\b B(t\pm \tau/2)$, and $\gamma=1/\sqrt{1-\beta^2}$. The tensor $p^{+}_{i}p^{-}_{k}$ has a quite complicated
structure. If the distribution of the particles is isotropic one can simplify $p^{+}_{i}p^{-}_{k}$  by averaging
directions of the velocity $\b \beta$ over the azimuthal angle with respect to the direction of observation 
$\b n$. The tensor can be further simplified taking into account that the angle between $\b\beta$ and $\b n$ is small ($\theta\sim 1/\gamma$) and $\beta\approx 1-1/2\gamma^2$. After tedious calculations, we obtain

\begin{eqnarray}
\langle p^{+}_{i}p^{-}_{k}\rangle=\frac{1}{8\gamma^4}T_{ik\mu\nu}B^{+}_{\mu}B^{-}_{\nu},
\end{eqnarray}
where the fourth-rank tensor $T_{ik\mu\nu}$ is
\begin{eqnarray}
T_{ik\mu\nu}=(2+\chi^2)\delta^{\perp}_{ik}\delta^{\perp}_{\mu\nu}-
\chi^2\delta^{\perp}_{i\mu}\delta^{\perp}_{k\nu}-
(2-\chi^2)\delta^{\perp}_{i\nu}\delta^{\perp}_{k\mu}.
\end{eqnarray}
Here we have introduced the variable $\chi=(\theta\gamma)^2$, where $\theta$ is the angle between $\b\beta$
and $\b n$. The tensor $\delta^{\perp}_{ik}=\delta_{ik}-n_in_k$ is the two-dimensional Kronecker delta perpendicular
to the direction $\b n$. 

Taking into account relativistic effects we can write in these designations
\begin{equation}
\eta\approx\frac{1}{2\gamma^2}(1+\chi).
\end{equation}
and
\begin{eqnarray}\label{eq:powik1}
(P_{\b n \omega}(t))_{ik}=\frac{e^4\gamma^2}{2\pi^2m^2c^3}\frac{1}{(1+\chi)^4}\times\\\nonumber
\int\limits_{-\infty}^{\infty}T_{ik\mu\nu}B^{+}_{\mu}B^{-}_{\nu}e^{i\omega\eta\tau}d\tau.
\end{eqnarray}

To continue the further calculations one should average Eq.~(\ref{eq:powik1}) over all possible configurations of the magnetic field. This leads to appearance of the correlation function
\begin{equation}\label{eq:kdef}
K_{\mu\nu}(\b r_1,t_1; \b r_2,t_2)=\langle B_{\mu}(\b r_1,t_1)B_{\nu}(\b r_2,t_2) \rangle,
\end{equation}
where $\b r_1=\b r+c\b \beta\tau/2$, $\b r_2=\b r-c\b \beta\tau/2$, $t_1=t+\tau/2$, and $t_2=t-\tau/2$.
We consider the case of the two-dimensional stochastic magnetic field with directions parallel to a plane.
Let us introduce a vector $\b s$ which is the normal unit vector to the plane. Then the correlation function should
satisfy the following conditions
\begin{equation}\label{eq:kpers}
K_{\mu\nu}s_{\mu}=0, \quad  K_{\mu\nu}s_{\nu}=0.
\end{equation}
Following to the procedure described in Ref.~\cite{Kelner2013}, the correlation function of the
statistically homogeneous and stationary field could be presented in the form of a Fourier integral:
\begin{equation}\label{eq:kexp}
K_{\mu\nu}(\b r,t)=\int \widetilde{K}_{\mu\nu}(\b q,\varkappa)e^{i(\b q\b r-\varkappa t)}
\frac{d^3q}{(2\pi)^3}\frac{d\varkappa}{2\pi},
\end{equation}
where $\b r=\b r_1-\b r_2=c\b \beta \tau$ and $t=t_1-t_2=\tau$.
Because of the condition $\nabla \b B=0$, the Fourier components of the correlation function obey the
transversality conditions:
\begin{equation}\label{eq:kperq}
\widetilde{K}_{\mu\nu}q_{\mu}=0, \quad  \widetilde{K}_{\mu\nu}q_{\nu}=0.
\end{equation}
Taking into account the conditions of Eq.~(\ref{eq:kpers}) and Eq.~(\ref{eq:kperq}), the most general form
of the correlation function is
\begin{eqnarray}\label{eq:kfform}
\widetilde{K}_{\mu\nu}=\langle \b B^2\rangle\tilde{\Psi}(|\b q|, \b s\b q,\varkappa)
\hat{K}_{\mu\nu}
\end{eqnarray}
where
\begin{equation}
\hat{K}_{\mu\nu}=(\delta_{\mu\nu}-\hat{q}_{\mu}\hat{q}_{\nu})-
\frac{1}{1-(\b s\hat{\b q})^2}(s_{\mu}-(\b s\hat{\b q})\hat{q}_{\mu})(s_{\nu}-(\b s\hat{\b q})\hat{q}_{\nu}),
\end{equation}
and $\hat{\b q}=\b q/q$.
Below we will consider the case of stationary magnetic fields with the function $\tilde{\Psi}$ which does not depend on
the scalar product $\b s\b q$:
\begin{equation}\label{eq:psit}
\tilde{\Psi}=\Psi(q)2\pi\delta(\varkappa).
\end{equation}
Here $\delta(\varkappa)$ is the Dirac delta function, and the function $\Psi$ is normalised such that
\begin{equation}\label{eq:psinorm}
\int \Psi(q)\frac{d^3 q}{(2\pi)^3}=\frac{1}{2\pi^2}\int\limits_{0}^{\infty} \Psi(q)q^2 dq=1.
\end{equation} 

Combining Eqs.~(\ref{eq:powik1},\ref{eq:kdef},\ref{eq:kexp},\ref{eq:kfform}) we obtain
\begin{eqnarray}\label{eq:powik2}
(\langle P_{\b n \omega}(t)\rangle)_{ik}=\frac{e^4\langle \b B^2\rangle\gamma^2}{2\pi^2m^2c^3}
\frac{1}{(1+\chi)^4}\times\\\nonumber
\int \Psi(q)T_{ik\mu\nu}\hat{K}_{\mu\nu}
\int\limits_{-\infty}^{\infty}e^{i(c\b q\b n+\omega\eta)\tau}d\tau \frac{d^3q}{(2\pi)^3}.
\end{eqnarray}
Here the integration over $\varkappa$ has been cancelled using $\delta(\varkappa)$ in Eq.~(\ref{eq:psit})
and the velocity $\b \beta$ in the exponent has been substituted by $\b n$.

To find the total radiation power emitted by an isotropic distribution of the particles in the direction $\b n$, let us integrate Eq.~(\ref{eq:powik2}) over directions of velocity $\b \beta$. Because the angle
between $\beta$ and $\b n$ is small, one can adopt $d\Omega=2\pi\theta d\theta$. It is convenient
to perform this integration using variable $\chi=(\theta\gamma)^2$ with limits of integration from zero
to infinity:
\begin{eqnarray}\label{eq:powik3}
(\bar{P}_{\b n \omega}(t))_{ik}=\frac{e^4\langle \b B^2\rangle}{m^2c^3}
\int\frac{d^3q}{(2\pi)^3}\times\\\nonumber
\int\limits_{0}^{\infty}\frac{d\chi}{(1+\chi)^4} \delta(c\b q\b n+\omega\eta)\Psi(q)T_{ik\mu\nu}\hat{K}_{\mu\nu}.
\end{eqnarray}

This integral is calculated using the delta function $\delta(cq\hat{\b q}\b n+\omega(1+\chi)/2\gamma^2)$ appearing after
the integration over $\tau$ in Eq.~(\ref{eq:powik2}):
\begin{eqnarray}\label{eq:powik4}
(\bar{P}_{\b n \omega}(t))_{ik}=\frac{e^4\langle \b B^2\rangle}
{m^2c^3}\frac{2\gamma^2}{\omega}\times\hspace{3cm}\\\nonumber
\frac{1}{(2\pi)^3}\int\limits_{\omega/2\gamma^2c}^{\infty}dq q^2 \Psi(q)\int\limits_{-1}^{-\xi}dx\int\limits_{0}^{2\pi} d\phi\frac{T_{ik\mu\nu}\hat{K}_{\mu\nu}}{(1+\chi_0)^4},
\end{eqnarray}
where 
\begin{equation}\label{eq:xisub}
\chi_0=-1-\frac{x}{\xi},\quad \xi=\frac{\omega}{2\gamma^2cq}.
\end{equation}
Here $x=\hat{\b q}\b n$ is the cosine of the angle between $\b q$ and $\b n$, $\phi$ is the azimuthal
angle of $\b q$ relative to the direction of $\b n$, and the tensor $T_{ik\mu\nu}=T_{ik\mu\nu}(\chi_0)$.

Taking into account that $\hat{K}_{\mu\nu}$ is symmetric tensor, the tensor part of the integrand is  simplified to
\begin{equation}\label{eq:tk}
T_{ik\mu\nu}\hat{K}_{\mu\nu}=(2+\chi^2_0)\delta^{\perp}_{ik}\hat{K}^{\perp}_{\mu\mu}-2 \hat{K}^{\perp}_{ik},
\end{equation}
where $\hat{K}^{\perp}_{ik}$ is a two-dimensional tensor perpendicular to the direction $\b n$; it is expressed as
\begin{equation}
\hat{K}^{\perp}_{ik}=\hat{K}_{\mu\nu}(\delta_{i\mu}-n_in_{\mu})(\delta_{k\nu}-n_kn_{\nu}).
\end{equation}

Let us introduce two-dimensional projection of the vector $\b s$ on the plane perpendicular to the direction
$\b n$
\begin{equation}\label{eq:detss}
\b s'=\frac{\b s - (\b s\b n)\b n}{|\b s - (\b s\b n)\b n|}.
\end{equation}
Then the tensor $\hat{K}^{\perp}_{ik}$ integrated over azimuthal angle $\phi$ should have the following structure
\begin{equation}
\int\limits_{0}^{2\pi}\hat{K}^{\perp}_{ik}d\phi=K_1\delta^{\perp}_{ik}+K_2s'_is'_k.
\end{equation}
The trace of the tensor and contraction with tensor $s'_is'_k$ results in two algebraic equations.
The solution of these equations is
\begin{eqnarray}
K_1=W_1-W_2,\\\nonumber
K_2=2W_2-W_1,
\end{eqnarray}
where
\begin{eqnarray}
W_1=\frac{1}{2\pi}\int\limits_{0}^{2\pi}\hat{K}^{\perp}_{\mu\mu} d\phi, \hspace{0.8cm} \\\nonumber
W_2=\frac{1}{2\pi}\int\limits_{0}^{2\pi}\hat{K}^{\perp}_{\mu\nu}s'_{\mu}s'_{\nu} d\phi.
\end{eqnarray}
The integration over the azimuthal angle gives
\begin{eqnarray}
W_1=\frac{|x+\sigma'|+|x-\sigma'|}{2},\hspace{2 cm}\\\nonumber
W_2=\frac{\sigma'^2}{1-\sigma'^2}\left(1-\frac{|x+\sigma'|+|x-\sigma'|}{2}\right),
\end{eqnarray}
where $\sigma'=\b s\b n$.
Thus the integration of Eq.~(\ref{eq:tk}) over the azimuthal angle $\phi$ gives
\begin{eqnarray}
\int\limits_{0}^{2\pi}T_{ik\mu\nu}\hat{K}_{\mu\nu} d\phi=\hspace{4 cm}\\\nonumber
W_1(1+\chi^2_0)\delta^{\perp}_{ik}-\left(W_1-2W_2\right)
(\delta^{\perp}_{ik}-2s'_is'_k).
\end{eqnarray}
Eq.~(\ref{eq:powik4}) can be rewritten as
\begin{eqnarray}\label{eq:powik5}
(\bar{P}_{\b n \omega}(t))_{ik}=\frac{e^4\langle \b B^2\rangle}{m^2c^4}
\left(\frac{\omega}{2\gamma^2c}\right)^2 \times\hspace{3cm}\\\nonumber
\frac{1}{(2\pi)^2}\int\limits_{0}^{1}\frac{d\xi}{\xi^3}\Psi\left(\frac{\omega}{2\gamma^2c\xi}\right) \left(F_1\delta^{\perp}_{ik}-F_2(\delta^{\perp}_{ik}-2s'_is'_k)\right),
\end{eqnarray}
where
\begin{eqnarray}
F_1=\xi^3\int\limits_{\xi}^{1}\frac{dx}{x^4}
\left(2-2\frac{x}{\xi}+\frac{x^2}{\xi^2}\right)W_1,\\\nonumber
F_2=\xi^3\int\limits_{\xi}^{1}\frac{dx}{x^4}
\left(W_1-2W_2\right).\hspace{1.2cm}
\end{eqnarray}
Here we have changed the integration variable from $q$ to $\xi$ using Eq.~(\ref{eq:xisub}) and
expressed $\chi_0$ through $\xi$. Also, for convenience of notations the interchange $x\rightarrow-x$ has been
applied. After rather simple but tedious calculations we obtain
\begin{eqnarray}\label{eq:f1}
F_1=\left(\frac{2\sigma}{3}+
\left(2-\frac{1}{\sigma}\right)\xi^2+\left(\frac{1}{3\sigma^2}-1\right)\xi^3\right.\hspace{2cm}
\\\nonumber
\left.-\xi(1+\ln\sigma)\right)
\Theta(\sigma-\xi)+
\xi(2\xi-\xi^2-1-\ln\xi)\Theta(\xi-\sigma),
\end{eqnarray}

\begin{eqnarray}\label{eq:f2}
F_2=\frac{1}{3}\left(\frac{1-\sigma}{1+\sigma}\right)\left(\frac{(1+\sigma)^2}{2\sigma^2}\xi^3+\sigma\right)\Theta(\sigma-\xi)+\hspace{1.3cm}\\\nonumber 
\frac{1}{1-\sigma^2}\left(\frac{1+\sigma^2}{2}(1-\xi^2)\xi-
\frac{2\sigma^2}{3}(1-\xi^3)\right)\Theta(\xi-\sigma),
\end{eqnarray}
where $\Theta$ is the Heaviside step function, $\sigma=|\sigma'|$.
Finally we can write the result as
\begin{equation}\label{eq:JitTensor}
(\bar{P}_{\b n \omega}(t))_{ik}=
\frac{e^4\langle \b B^2\rangle}{m^2c^4}\left(I_j\delta_{ik}-Q_j(\delta_{ik}-2s'_is'_k)\right),
\end{equation}
where we suppose that $\delta_{ik}$ is the two-dimensional Kronecker delta, and
\begin{eqnarray}\label{eq:QRJit}
I_j=\left(\frac{\omega}{2\gamma^2c}\right)^2\frac{1}{(2\pi)^2}\int\limits_{0}^{1}\frac{d\xi}{\xi^3}\Psi\left(\frac{\omega}{2\gamma^2c\xi}\right)F_1(\xi,\sigma)\\\nonumber
Q_j=\left(\frac{\omega}{2\gamma^2c}\right)^2\frac{1}{(2\pi)^2}\int\limits_{0}^{1}\frac{d\xi}{\xi^3}\Psi\left(\frac{\omega}{2\gamma^2c\xi}\right)F_2(\xi,\sigma).
\end{eqnarray}
Here the functions $F_1$ and $F_2$ are given by Eqs.~(\ref{eq:f1}), (\ref{eq:f2}). The functions $I_j$ and $Q_j$ correspond up to the common prefactor to the Stokes parameters
$I$ and $Q$ as they are usually determined. The lower index $j$ refers to the jitter radiation. The results written in this form are convenient because the degree of polarization can be immediately written as:
\begin{equation}\label{eq:QR}
\Pi_j=\frac{Q_j}{I_j}.
\end{equation} 
Indeed, the intensity observed in the direction $\b e$ is
\begin{equation}
I(\theta)=(\bar{P}_{\b n \omega}(t))_{ik}\b e_i\b e_k=\frac{e^4\langle \b B^2\rangle}{m^2c^4}
\left(I_j-Q_j(1-2\cos^2\phi_p)\right),
\end{equation} 
where $\phi_p$ is the angle between two two-dimensional vectors $\b s'$ and $\b e$. As 
$I_{min}=I(90^\circ)\sim I_j-Q_j$ and $I_{max}=I(0^\circ)\sim I_j+Q_j$, from the definition
$\Pi=(I_{max}-I_{min})/(I_{max}+I_{min})$ we obtain Eq.~(\ref{eq:QR}).

Let us find the total radiation by integrating the power spectrum $(\bar{P}_{\b n \omega}(t))_{ik}$ over the frequency $\omega$. Taking into account the normalization condition given by Eq.~(\ref{eq:psinorm}), we obtain
\begin{eqnarray}
(\bar{P}_{\b n}(t))_{ik}=\frac{e^4\langle \b B^2\rangle}{m^2c^3}\gamma^2
\int\limits_{0}^{1}d\xi\left(F_1\delta_{ik}-F_2(\delta_{ik}-2s'_is'_k)\right).
\end{eqnarray}
The calculations give
\begin{eqnarray}
I'_j=\int\limits_{0}^{1}d\xi F_1=\frac{1+\sigma^2}{6},\\\nonumber
Q'_j=\int\limits_{0}^{1}d\xi F_2=\frac{1-\sigma^2}{8}.
\end{eqnarray}
One can show that the polarisation of the total radiation is expressed as
\begin{equation}\label{eq:intpol}
\Pi'_j=\frac{Q'_j}{I'_j}=\frac{3}{4}\left(\frac{1-\sigma^2}{1+\sigma^2}\right).
\end{equation}
Eq.~(\ref{eq:intpol}) represents a general expression for the degree of polarization of the total intensity of
ultrarelativistic particles in the magnetic field of given geometry. Below we show that this expression also appears in the case of the synchrotron radiation.

The total intensity averaged over directions of $\b n$ ($d\Omega_{\b n}/4\pi$) is
\begin{equation}\label{eq:AvInten}
(\bar{P}(t))_{ii}=\frac{4}{9}\frac{e^4\langle \b B^2\rangle}{m^2c^3}\gamma^2.
\end{equation}
It coincides with the radiation intensity in the totally chaotic magnetic field (without a preferred direction).

Averaging of the power spectrum over directions of $\b n$ leads to
\begin{equation}
(\bar{P}_{\omega}(t))_{ii}=\frac{e^4\langle \b B^2\rangle}{6\pi^2m^2c^3}
\int\limits_{\omega/2c\gamma^2}^{\infty}u(y)\Psi(q)qdq,
\end{equation}
where $y=2qc\gamma^2/\omega$, and
\begin{equation}
u(y)=1+\frac{3}{y^2}-\frac{4}{y^3}-\frac{3\ln y}{y^2}.
\end{equation}
The expression coincides with the formula obtained in Ref.~\cite{Kelner2013}.

\section{Polarization of synchrotron radiation}
The polarization of synchrotron radiation in the case of homogeneous magnetic field is described
by the matrix \cite{Rybicki1986}:
\begin{equation}\label{eq:SpSinMat}
P_{ik}=\frac{\sqrt{3}}{4\pi}\frac{e^2\sin\vartheta}{R_L}
\begin{pmatrix}
  F(x)+G(x) & 0 \\
  0 & F(x)-G(x)
 \end{pmatrix},
\end{equation}
where 
\begin{equation}
x=\frac{\omega}{\omega_c\sin\vartheta},\quad \omega_c=\frac{3}{2}\gamma^2\omega_{B},\quad R_L=\frac{c}{\omega_B}.
\end{equation}
Here $\omega_B=\frac{eB}{mc}$ is the cyclotron frequency of electrons in the magnetic field with the strength
$B$, $\vartheta$ and $\gamma$ are the pitch angle and Lorentz factor of the radiating electron.
The function $F(x)$ and $G(x)$ are expressed through the modified Bessel functions $K_{5/3}(x)$ and $K_{2/3}(x)$ in the following way:
\begin{equation}
F(x)=x\int\limits_{x}^{\infty}K_{5/3}(t)dt,\quad G(x)=xK_{2/3}(x).
\end{equation}
The power spectrum of synchrotron radiation is obtained as the trace of the matrix in Eq.~(\ref{eq:SpSinMat}):
\begin{equation}
P_{\omega}=P_{ii}=\frac{\sqrt{3}}{2\pi}\frac{e^2}{R_L}\sin\vartheta F(x).
\end{equation}
The polarisation matrix in Eq.~(\ref{eq:SpSinMat}) is written in the plane
perpendicular to the direction of observation $\b n$, in a particular system of coordinates where $x$ and $y$ axes
are directed along acceleration $\b w$ and along the direction $\b n\times\b w$ perpendicular to it, respectively.
Let us rewrite Eq.~(\ref{eq:SpSinMat}) for an arbitrary reference coordinate system in the tensor form:
\begin{equation}
P_{ik}=\frac{\sqrt{3}}{4\pi}\frac{e^2}{R_L \sin\vartheta} \rho_{ik},
\end{equation}
where
\begin{eqnarray}
\rho_{ik}=\sin^2\vartheta(\left(F(x)+G(x)\right)\epsilon_{i}\epsilon_{k}\\\nonumber
+\left(F(x)-G(x)\right)(\b n\times\b \epsilon)_{i}(\b n\times\b \epsilon)_{k}).
\end{eqnarray}
Here $\b\epsilon=\b w/w$ is the unit vector in the direction of acceleration $\b w=\frac{eB}{m\gamma}(\b\beta\times\b b)$, where $\b b$ is the unit vector in the direction of magnetic field.
In this expression the velocity $\b\beta$ can be substituted by $\b n$. Thus we obtain 
\begin{equation}
\b \epsilon=\frac{\b n\times \b b}{\sin\vartheta},
\end{equation}
and 
\begin{eqnarray}\label{eq:rhos}
\rho_{ik}=\left(F(x)+G(x)\right)(\b n\times\b b)_{i}(\b n\times\b b)_{k}\\\nonumber
+\left(F(x)-G(x)\right)(\b n\times(\b n\times\b b))_{i}(\b n\times(\b n\times\b b))_{k}.
\end{eqnarray}
Let us assume that magnetic field has a slab structure, i.e. the directions of the magnetic field are
parallel to a plane. Designating the normal vector to the plane as $\b s$, we can always choose two 
perpendicular to each other vectors $\b e_{1}$ and $\b e_{2}$ lying in the plane in a way that vectors
$\b e_{1}$, $\b e_{2}$, and $\b s$ constitute a right-handed coordinate system. Then the vectors $\b n$ and $\b b$ are expressed as
\begin{eqnarray}\label{eq:ne1s}
\b n=\sin\theta\,\b e_1+\cos\theta \b s,\quad\\\label{eq:be1e2}
\b b=\cos\phi_B\b e_1+\sin\phi_B\b e_2.
\end{eqnarray}
It follows from Eq.~(\ref{eq:ne1s}) that
\begin{eqnarray}\label{eq:e1e2}
\b e_1=\frac{1}{\sin\theta}(\b n-(\b n\b s)\b s),\qquad\\\nonumber
\b e_2=\b s\times\b e_1=\frac{1}{\sin\theta}(\b s\times \b n).
\end{eqnarray} 
The substitution of Eq.~(\ref{eq:e1e2}) to Eq.~(\ref{eq:be1e2}) leads to
\begin{equation}
\b b=\frac{1}{\sin\theta}\left(\cos\phi_B(\b n-(\b n\b s)\b s)+\sin\phi_B (\b s\times \b n)\right).
\end{equation}
Then we have
\begin{equation}\label{eq:ntb}
\b n\times\b b=\frac{1}{\sin\theta}\left(-\cos\phi_B(\b n\b s)(\b n\times\b s)+
\sin\phi_B (\b s-(\b n\b s)\b n)\right).
\end{equation}
Notice that $\b n\times\b b$ is perpendicular to $\b n$ and has only two non-zero components in the
system of coordinates with one of the axis directed along $\b n$. Indeed the vector
$\b s-(\b n\b s)\b n$  is the projection of vector $\b s$ on the plane perpendicular to $\b n$ and
can be rewritten as
\begin{equation}
(\b s-(\b n\b s)\b n)_i=\sin\theta s'_i,
\end{equation}
where $\b s'$ is determined in Eq.~(\ref{eq:detss}).
Further we consider $\b s'$ as a two-dimensional vector. Analogously, the vector $\b n\times\b s$ could be written as
\begin{equation}
(\b n\times\b s)_i=-\sin\theta \varepsilon_{ij} s'_j,
\end{equation}
where $\varepsilon_{ij}$ is a two-dimensional antisymmetric symbol. Thus, in the plane perpendicular
to $\b n$ we can write
\begin{equation}
(\b n\times\b b)_i=\cos\phi_B(\b n\b s)\varepsilon_{ij} s'_j+\sin\phi_B  s'_i,\quad i=1,2.
\end{equation}
Introducing $M=\cos\phi_B(\b n\b s)$ and $N=\sin\phi_B$, we have
\begin{equation}
(\b n\times\b b)_i=M\varepsilon_{ij} s'_j+N  s'_i.
\end{equation}
Then the first tensor in Eq.~(\ref{eq:rhos}) can be written as
\begin{eqnarray}
(\b n\times\b b)_i(\b n\times\b b)_k=M^2(\delta_{ik}-s'_is'_k)+\\\nonumber
MN(s'_i\varepsilon_{k\mu}s'_{\mu}+\varepsilon_{i\nu}s'_{\nu}s'_{k})+N^2s'_is'_k,
\end{eqnarray}
where $\delta_{ik}$ is the two-dimensional Kronecker delta.

Taking the vector product of $\b n$ and $\b n\times\b b$ from Eq.~(\ref{eq:ntb}) we obtain
\begin{eqnarray}
\b n\times(\b n\times\b b)=\frac{1}{\sin\theta}\left(\cos\phi_B(\b n\b s)(\b s-(\b n\b s)\b n)\right.+
\\\nonumber
\left.\sin\phi_B (\b n\times\b s)\right).
\end{eqnarray}   
Analogously, this vector can be written in two-dimensional form:
\begin{eqnarray}
(\b n\times(\b n\times\b b))_i=Ms'_i-N\varepsilon_{ij} s'_j.
\end{eqnarray}
The second tensor in Eq.~(\ref{eq:rhos}) has the following form
\begin{eqnarray}
(\b n\times(\b n\times\b b))_i(\b n\times(\b n\times\b b))_k=M^2 s'_is'_k-\\\nonumber
MN(s'_i\varepsilon_{k\mu}s'_{\mu}+\varepsilon_{i\nu}s'_{\nu}s'_{k})+N^2(\delta_{ik}-s'_is'_k).
\end{eqnarray}

The direction of the magnetic field enters also in the argument of the functions $F(x)$ and $G(x)$
through the sine of the pitch angle $\sin\vartheta$, which can be expressed as 
\begin{equation}
\sin\vartheta=|\b n\times\b b|=\sqrt{\sin^2\phi_b+(\b n\b s)^2\cos^2\phi_B}.
\end{equation}
Taking in mind that at averaging over the angle $\phi_B$ the expressions with the factor $MN=\sin\phi_B\cos\phi_B(\b n\b s)$ drop out, the tensor $\langle \rho_{ik}\rangle$ in Eq.~(\ref{eq:rhos}) can be written as
\begin{eqnarray}
\langle \rho_{ik}\rangle=\langle\left(F(x)+G(x)\right)(M^2(\delta_{ik}-s'_is'_k)+N^2s'_is'_k)\\\nonumber
+\left(F(x)-G(x)\right)(M^2s'_is'_k+N^2(\delta_{ik}-s'_is'_k))\rangle,
\end{eqnarray}
or in the more convenient form
\begin{eqnarray}
\langle \rho_{ik}\rangle=\langle F(x)\sin^2\vartheta\delta_{ik}\\\nonumber
-G(x)(2\sin^2\phi_B-\sin^2\vartheta)(\delta_{ik}-2s'_is'_k)\rangle.
\end{eqnarray}
Finally we can write
\begin{eqnarray}\label{eq:ppolar}
\langle P_{ik}\rangle=\frac{\sqrt{3}}{4\pi}\frac{e^2}{R_L}(I_s\delta_{ik}-Q_s(\delta_{ik}-2s'_is'_k)),
\end{eqnarray}
where
\begin{eqnarray}\label{eq:pQR}
I_s\left(\frac{\omega}{\omega_c}\right)=\frac{2}{\pi}\int\limits_{0}^{\frac{\pi}{2}}d\phi_B \chi F\left(\frac{\omega}{\omega_c\chi}\right),\qquad\qquad\qquad\\\nonumber
Q_s\left(\frac{\omega}{\omega_c}\right)=\frac{2}{\pi}\int\limits_{0}^{\frac{\pi}{2}}d\phi_B \left(2\frac{\chi^2-\sigma^2}{1-\sigma^2}-\chi^2\right)\frac{1}{\chi}G\left(\frac{\omega}{\omega_c\chi}\right).
\end{eqnarray}
Here 
\begin{eqnarray}
\chi=\sin\vartheta=\sqrt{1-(1-\sigma^2)\cos^2\phi_B}
\end{eqnarray}
where  $\sigma=\b n \b s$. Here, as before, the functions $I_s$ and $Q_s$ correspond up to the 
prefactor to the Stokes parameters $I$ and $Q$. The lower index $s$ refers to the synchrotron radiation. 

If the magnetic field is distributed over the strength with the distribution function $w(B)$,
we should average the power spectrum in Eq.~(\ref{eq:ppolar}) 
\begin{equation}
\overline{P_{ik}}=\int\limits_{0}^{\infty}\langle P_{ik}\rangle w(B)dB.
\end{equation}
Notice that this operation does not change the tensor structure of $\langle P_{ik}\rangle$.
The total power spectrum can be found by taking the trace of $\langle P_{ik}\rangle$:
\begin{equation}
\langle P_{ii}\rangle=\frac{\sqrt{3}}{2\pi}\frac{e^2}{R_L}I_s\left(\frac{\omega}{\omega_c}\right)
\end{equation}
The power spectrum in some particular direction $\b e$ in the plane perpendicular to $\b n$ is
\begin{equation}
\langle P_{ik}\rangle e_ie_k=\frac{\sqrt{3}}{4\pi}\frac{e^2}{R_L}(I_s-Q_s(1-2(\b s'\b e)^2)),
\end{equation}
i.e. it depends on the scalar product between projection of the normal vector $\b s$ and the
direction of the polarizer $e$. Note that the polarization tensor has the same structure as for the case of the jitter radiation. Thus, the degree of polarization of the synchrotron radiation in the slab geometry is
\begin{equation}
\Pi_s=\frac{Q_s}{I_s}.
\end{equation}

The integration of the $\langle P_{ik}\rangle$ over the frequency in Eq.~(\ref{eq:pQR}) leads to
\begin{eqnarray}
I_s'=\int\limits_{0}^{\infty}d\omega I_s\left(\frac{\omega}{\omega_c}\right)=\omega_c\frac{8\pi}{27}\sqrt{3}\frac{1+\sigma^2}{2},\\\nonumber
Q_s'=\int\limits_{0}^{\infty}d\omega Q_s\left(\frac{\omega}{\omega_c}\right)=\omega_c\frac{2\pi}{9}\sqrt{3}\frac{1-\sigma^2}{2}.
\end{eqnarray}
Then the degree of polarization of the total radiation is
\begin{equation}
\Pi'_s=\frac{Q_s'}{I_s'}=\frac{3}{4}\left(\frac{1-\sigma^2}{1+\sigma^2}\right),
\end{equation}
which coincides with Eq.~(\ref{eq:intpol}).

\section{Results}
Now we analyse the results obtained in the previous sections. In particular, we use Eqs.~(\ref{eq:JitTensor}-\ref{eq:QRJit}) for the jitter radiation and Eqs.~(\ref{eq:ppolar}-\ref{eq:pQR}) for the
synchrotron radiation.

\subsection{Magnetic field distributions}
The expressions obtained in the  previous sections do not specify the exact form of the turbulent spectrum or the distribution over the strength. Below we describe the distributions which are used for the presentation of the final results.

In the regime of synchrotron radiation the spatial correlation of the magnetic fields has no impact on the radiation spectrum. Therefore the variation of the magnetic field can be considered as a set of homogeneous magnetic fields with a certain distribution over the strength and the direction without an introduction of the turbulent spectrum.
The calculations leading to Eq.~(\ref{eq:pQR}) have been conducted for the isotropic in the plane magnetic field with the constant strength. Assuming that the variations of the magnetic field strength are distributed isotropically and homogeneously along the plane, one can describe the large-scale turbulence via the distribution over the magnetic field strength:
\begin{equation}
w(B)dB=h_{n}(b)dB/B_0,
\end{equation}
where $B_0\equiv\sqrt{\langle\b B^2\rangle}$ and $b=B/B_0$. In the same manner as in Ref.~\cite{Kelner2013} we
consider the following three types of distributions:
\begin{eqnarray}\label{eq:StrDist}
h_0(b)=\delta(b-1),\\\nonumber
h_1(b)=\frac{3\sqrt{6}}{\sqrt{\pi}}b^2e^{-3b^2/2},\\\nonumber
h_2(b)=\frac{32b^2}{\pi(1+b^2)^4}.
\end{eqnarray}
Here $h_0$ describes homogeneous magnetic field, $h_1$ is the Gauss type distribution with narrow dispersion, and $h_2$ is the distribution of power-law type with wide dispersion around $B_0$.
For all distributions we have:
\begin{equation}\label{eq:DistVar}
\int\limits_{0}^{\infty}h_{n}(b)db=\int\limits_{0}^{\infty}b^2h_{n}(b)db=1.
\end{equation}

Unlike the synchrotron radiation, the jitter radiation depends on the spacial correlations of the magnetic field, and the turbulence spectrum determines the form of the radiation spectrum. Eq.~(\ref{eq:QRJit}) is derived for the 
homogeneous isotropic turbulence in the plane. In this case the spectrum of turbulence is described by the 
scalar function $\Psi$. It is convenient to present this function in the form
\begin{equation}\label{eq:JitTurbSpec}
\Psi(q)=\frac{A_{\alpha}\lambda^3}{(1+\lambda^2q^2)^{1+\alpha/2}},
\end{equation}
which provides convergence of the integrals in Eq.~(\ref{eq:QRJit}) at $\alpha>1$.

The normalization constant that satisfies the condition of Eq.~(\ref{eq:psinorm}) is
\begin{equation}
A_{\alpha}=8\pi^{3/2}\frac{\Gamma(1+\alpha/2)}{\Gamma((\alpha-1)/2)},
\end{equation}
where $\Gamma(z)$ is the gamma function. Note that the presence of the large-scale turbulence in the form
given by Eq.~(\ref{eq:StrDist}) does not change the results. 
Indeed, the variance of a small-scale turbulent magnetic field $\langle \b B^2\rangle$  enters linearly to the expressions for the jitter radiation. Averaging of $B^2$ over strength distributions gives, according to Eq.~(\ref{eq:DistVar}), $B^2$ again.
 
\subsection{Jitter radiation}

\begin{figure}
\begin{center}
\mbox{\includegraphics[width=0.5\textwidth,angle=0]{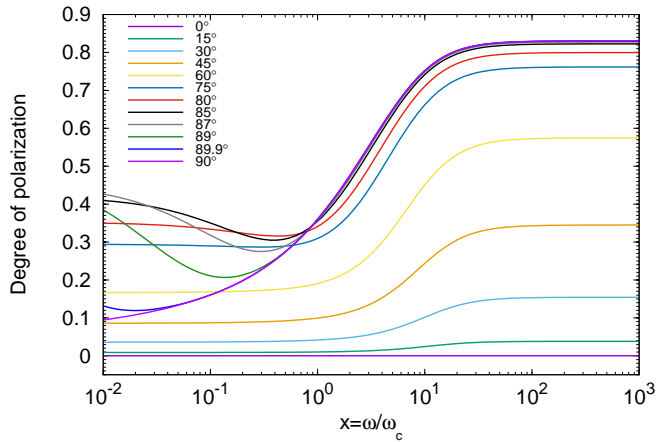}}
\caption{The of polarization of the jitter radiation as a function of frequency for different observation
angles. The curves correspond to $\lambda/R_L=0.1$ and the spectrum of turbulence with index $\alpha=5/3$. 
\label{fig:JitPolarization53}}
\end{center}
\end{figure}

\begin{figure}
\begin{center}
\mbox{\includegraphics[width=0.5\textwidth,angle=0]{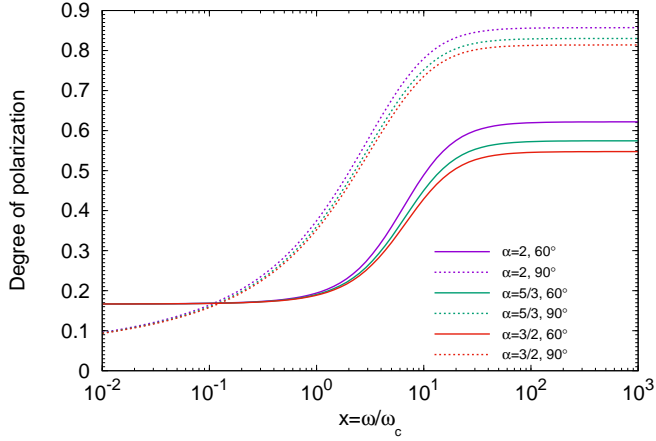}}
\caption{The degree of polarization of the jitter radiation for different spectra of turbulence with an index $\alpha$ calculated for the
characteristic coherence length $\lambda/R_L=0.1$, and two different observation angles $60^{\circ}$ and $90^{\circ}$. 
\label{fig:JitPolarization}}
\end{center}
\end{figure}

The results for the degree of polarization of jitter radiation based on Eqs.~(\ref{eq:QRJit},\ref{eq:QR}) are
shown in Figs.~\ref{fig:JitPolarization53}-\ref{fig:JitPolarization}. The curves are calculated for the case
of the turbulent spectrum given by Eq.~(\ref{eq:JitTurbSpec}) with correlation length $\lambda=0.1R_L$. The typical
behaviour of the polarization degree with frequency for different observation angles is shown in
Fig.~\ref{fig:JitPolarization53} for the spectral index of turbulence $\alpha=5/3$. The observation angle
is counted from the normal to the plane with chaotic magnetic field. 

If one observes the plane along the normal (with line of sight perpendicular to the plane), the radiation appears totally unpolarized. In this case, there is no preferential direction because the directions of the magnetic field are isotropically distributed and the fluxes of the radiation with different polarization are equal. However, when the observation is conducted at an angle to the normal, the isotropic picture is broken: while the angle between the line of sight and the directions of magnetic field perpendicular to the plane formed by line of sight and the normal remains $90^{\circ}$, the angle in other directions changes. Thus, the fluxes of the radiation with polarization in one direction would be dominant compared to the fluxes with polarization in other directions. One can see in Fig.~\ref{fig:JitPolarization53} that in accordance with this consideration the polarization generally increases with the angle. However, at small frequencies and at angles close to $90^{\circ}$ the opposite behaviour is seen.

The analysis of the results presented in Figs.~\ref{fig:JitPolarization53}-\ref{fig:JitPolarization} shows that
at angles greater and smaller than $\theta_{crit}\approx68.4^{\circ}$, the behaviour of the polarization with frequency is different. Namely, at the angles smaller than $\theta_{crit}$ the derivative
of the degree of polarization is positive at any frequency which means that the polarization is a strictly increasing function of frequency. At low and high frequencies the polarization grows slowly producing
some kind of plateaus. Between these plateaus one can observe sharp increase of the polarization located between
$\sim0.1\omega_{j}$ and $\sim\omega_{j}$ around the characteristic frequency of the jitter radiation $\omega_{j}=10\omega_{c}$. The transition region between the plateaus becomes sharper and starts at lower frequencies with the increase of the observation angel.

At the critical angle $\theta_{crit}$ the polarization changes the sign of the derivative at the point  $\omega_{turn}\approx3\cdot10^{-3}\omega_{c}$ becoming a decreasing function at small frequencies.
For the angle $75^{\circ}$ shown in Fig.~\ref{fig:JitPolarization53} this values is $\omega_{turn}\approx0.29\omega_{c}$. Because of the very slow variation, the change of trend at this frequency is barely seen. For larger angles the point of the minimum where the derivative changes its sign is more pronounced. At the angle $90^{\circ}$ this point shifts to zero. Thus, from the physical point of view, one can consider that at the observation angle $90^{\circ}$ the behaviour of the degree of polarization as function of frequency again changes to monotonically increasing function. One of the explanations of such an interesting behaviour of polarization with the observation angle at small frequencies could be that this is a mathematical artefact due to the delta-functional distribution of the directions of the magnetic field in three-dimensional space (all the directions lie in the plane).

One can show that the polarization degree converges to the constant limit at high frequencies. Indeed, at high frequencies the turbulent spectrum behaves as
\begin{equation}\label{eq:PsiHigh}
\Psi(q) \underset{q\rightarrow\infty}{\sim}  \frac{1}{q^{2+\alpha}}.
\end{equation}
Thus, the degree of polarization is
\begin{equation}
\Pi_j=\frac{\int\limits_{0}^{1}d\xi \xi^{\alpha-1}F_{2}(\xi,\sigma)}{\int\limits_{0}^{1}d\xi \xi^{\alpha-1}F_{1}(\xi,\sigma)}.
\end{equation}
After calculation of the integral we obtain the analytical expression for the maximum of the jitter polarization
which depends on the angle of observation ($\sigma=\cos \theta$) and the spectral index of the turbulence $\alpha$:
\begin{equation}\label{eq:jHighLimit}
\Pi_j=\frac{(1+\alpha)(2+\alpha)}{\alpha^2+3\alpha+4}
\left(1-\frac{2\sigma^2(1-\sigma^{1+\alpha})}{(1-\sigma^2)(\alpha+\sigma^{1+\alpha})}\right).
\end{equation}

\begin{figure}
\begin{center}
\mbox{\includegraphics[width=0.5\textwidth,angle=0]{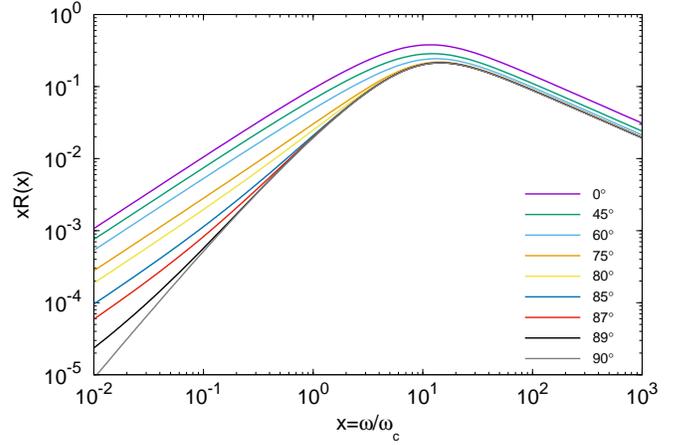}}
\caption{\label{fig:TotalPowerJitter53}The spectral energy distribution of the jitter radiation at different observation angles. The curve
is calculated for the case of the characteristic coherence length $\lambda/R_L=0.1$ and the turbulent spectrum index $\alpha=5/3$.}
\end{center}
\end{figure}

The value of the high-frequency plateau of the degree of polarization increases with the observation angle and spectral index. Specifically, at the angle $\theta=90^{\circ}$ the polarization
\begin{equation}
\Pi_j=\frac{\alpha^2+3\alpha+2}{\alpha^2+3\alpha+4}
\end{equation}  
is as high as $81\%$, $83\%$, and $86\%$ for the spectral indices $3/2$, $5/3$, $2$, respectively (see Fig.~\ref{fig:JitPolarization}). The polarization at the maximum of the spectral energy distribution
$\omega_{j}$ is $73\%$, $75\%$, and $78\%$.

Using Eqs.~(\ref{eq:QRJit}) and (\ref{eq:JitTurbSpec}) one can show that at low frequencies the degree of polarization as function of frequency also tends to the constant which is now independent on the spectral index of turbulence. After quite tedious calculations one can obtain that this constant is
\begin{equation}\label{eq:jLowLimit}
\Pi_j=\frac{1}{2}\left(\frac{1-\sigma}{1+\sigma}\right).
\end{equation}
At the observational angle $90^{\circ}$ this limit is reached at zero frequency. But at angles close to $90^{\circ}$ the polarization can attain the low frequency plateau at physically meaningful frequencies and can
be as high as almost $50\%$.

As it is clear from Eqs.~(\ref{eq:jHighLimit}) and (\ref{eq:jLowLimit}), and Fig.~(\ref{fig:JitPolarization}) that the degree of polarization does not depend on the spectral index of turbulence at low frequencies and increases with it at high frequencies. The higher value of the spectral index corresponds to the turbulence when more energy is concentrated at larger scales of turbulence. This means that the turbulence with smaller scale gives less polarized radiation at high frequencies.

\begin{figure}
\begin{center}
\mbox{\includegraphics[width=0.5\textwidth,angle=0]{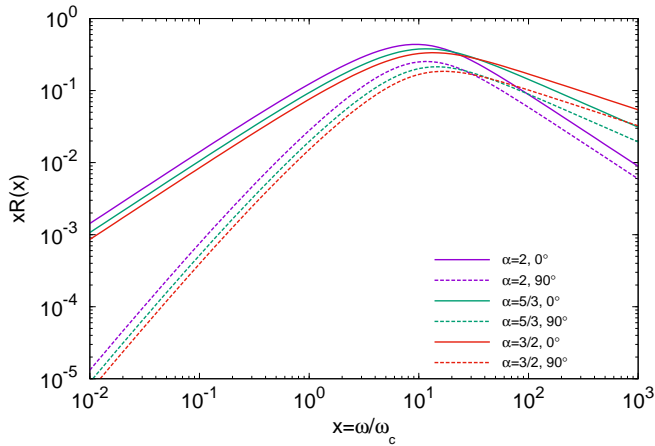}}
\caption{\label{fig:TotalPowerJitter} The spectral energy distribution of jitter radiation for different indices of the turbulent spectrum $\alpha$ and
the characteristic coherence length $\lambda/R_L=0.1$. Each case is shown for two observation angles $0^{\circ}$ and $90^{\circ}$.}
\end{center}
\end{figure}

The spectral power of jitter radiation also depends on the observation angle. It is convenient to plot the spectral energy distribution normalized to the total averaged intensity given by Eq.~(\ref{eq:AvInten}):
\begin{equation}
R(x)dx=P_{ii}(\omega)d\omega/\bar{P}_{ii}=2I_j(\omega)d\omega/\bar{P}_{ii}, \quad x=\omega/\omega_{c}.
\end{equation}
The result is presented in Fig.~\ref{fig:TotalPowerJitter53}. The form of the spectrum does not change at large
frequencies but slightly decreases with the angle. It can be shown that the high frequency slope is determined by the spectral index of turbulence. Indeed, in accordance with Eq.~(\ref{eq:PsiHigh}), Eq.~(\ref{eq:QRJit}) gives
\begin{equation}\label{eq:JHighPower}
I_j\underset{\omega\rightarrow\infty}{\sim}\frac{1}{\omega^{\alpha}}.
\end{equation}
At low frequencies, the form of the spectrum depends on the angle. In the limit of the small frequencies the 
behaviour of the function $R(x)$ is well described by the following expression:
\begin{equation}\label{eq:assymp}
R(x)\underset{x\rightarrow 0}{\sim}\frac{2\sigma}{3\alpha}-
\frac{3}{4}\frac{\sqrt{\pi}\,\Gamma\left((3+\alpha)/2\right)}{(1+\alpha)\Gamma\left(1+\alpha/2\right)}\delta_{j} (1+\ln \sigma)x,
\end{equation}
where a common constant is omitted and $\delta_{j}=\lambda/R_{L}$. From this expression one can see that spectral
power tends to a constant. It should be noted that at angles smaller than $\theta_{crit}=\arccos(1/e)\approx68.4^{\circ}$ the function tends to the constant from below, whereas at larger angles it tends to the same constant from above. At $\theta=90^{\circ}$ the function behaves as $R(x)\sim x$. The maximum of the spectral energy distribution is around $\omega_{j}$. Taking into account a prefactor omitted in Eq.~(\ref{eq:assymp}) one can show that at small frequencies the function $R(x)$ grows
with $\alpha$ as it can be seen from Fig.~\ref{fig:TotalPowerJitter}. 

\subsection{Synchrotron radiation}

\begin{figure}
\begin{center}
\mbox{\includegraphics[width=0.5\textwidth,angle=0]{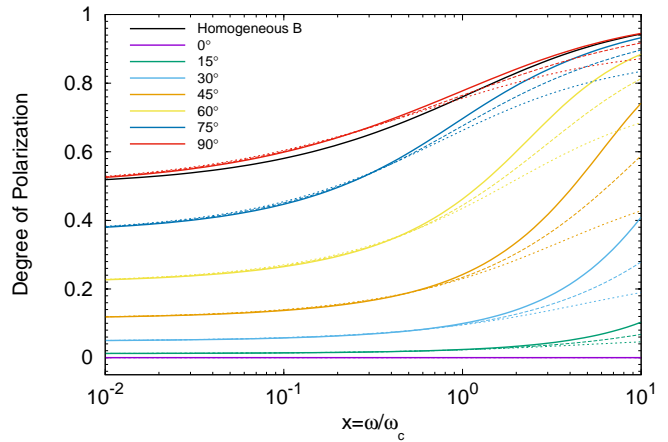}}
\caption{The degree of polarization of the synchrotron radiation as a function of frequency calculated for different observation angles. The curves corresponding to the magnetic field strength distributions $h_0$, $h_1$, and $h_2$ are shown by solid, dashed, and dotted lines, respectively. The curve indicated as 
'Homogeneous B' is the case of radiation of an electron with~$90^{\circ}$ pitch-angle in the homogeneous magnetic field.\label{fig:SynPolarization}}
\end{center}
\end{figure}

\begin{figure}
\begin{center}
\mbox{\includegraphics[width=0.5\textwidth,angle=0]{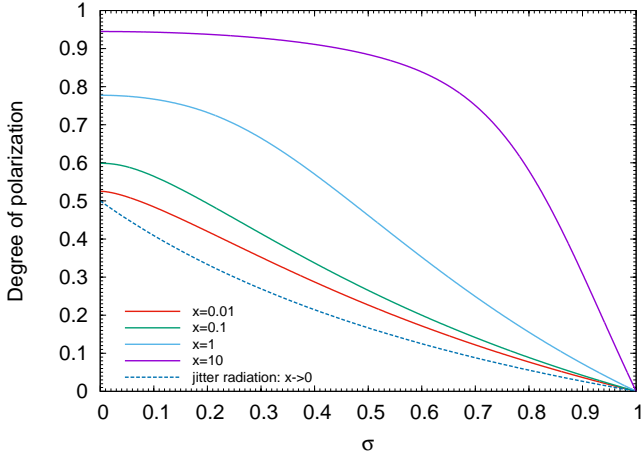}}
\caption{\label{fig:PolAngle}The degree of polarization of synchrotron radiation in the case of the magnetic field distribution $h_0$ as a function of the cosine of the observation angle $\sigma=\cos\theta$ calculated for different frequencies $x=\omega/\omega_{c}$. The dashed line corresponds to the low-frequency limit of the jitter radiation given by Eq.~(\ref{eq:jLowLimit})}.
\end{center}
\end{figure}

The degree of polarization of the synchrotron radiation is shown in Fig.~\ref{fig:SynPolarization}. For the same reason as in the case of the jitter radiation, the polarization is zero at the observation angle $0^{\circ}$ and increases with the observation angle. The curve indicated as 'Homogeneous B' presents the case of radiation of an electron with the pitch-angle $90^{\circ}$ in the homogeneous magnetic field. The polarization for the case of the distributions of the magnetic field strength $h_0$, $h_1$, and $h_2$ are shown by solid, dashed, and dotted curves, respectively. One can see that for all angles the polarization grows with frequency and tends to a constant at small frequencies. In the case of 'Homogeneous B' this constant equals $50\%$.
As expected, at large frequencies the polarization is smaller for broader distributions of the magnetic field strength.  Fig.~\ref{fig:PolAngle} shows the degree of polarization as a function of observation angle. At the characteristic frequency the polarization can be as high as $78\%$. Note that the polarization increases with the observation angle faster at high frequencies.

\begin{figure}
\begin{center}
\mbox{\includegraphics[width=0.5\textwidth,angle=0]{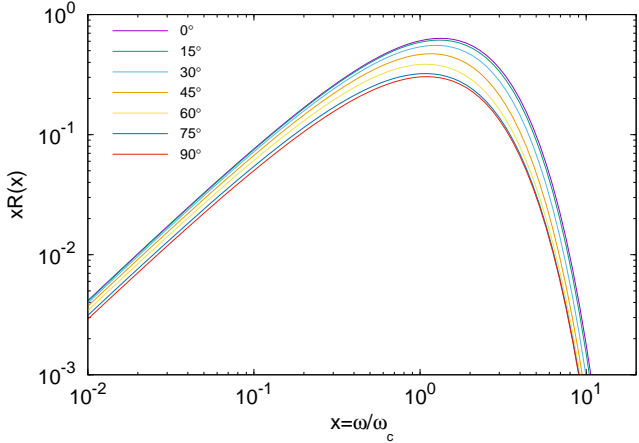}}
\caption{\label{fig:TotalPowerSyn}The spectral energy distribution of the synchrotron radiation for different observation angles.}
\end{center}
\end{figure}

\begin{figure}
\begin{center}
\mbox{\includegraphics[width=0.5\textwidth,angle=0]{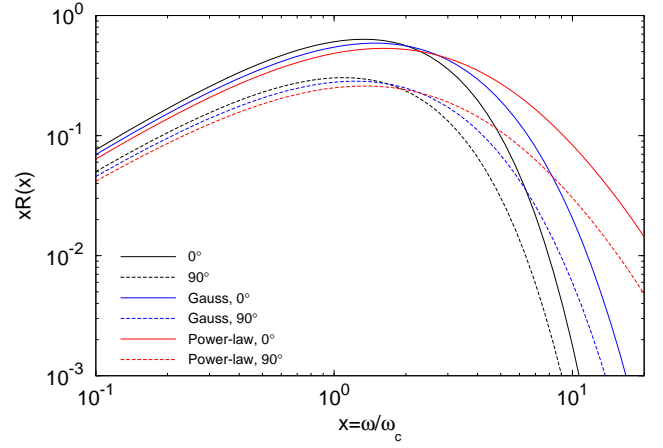}}
\caption{\label{fig:TotalPowerSynB}The spectral energy distribution of the synchrotron radiation calculated for the magnetic field strength distributions $h_0$,
$h_1$, and $h_2$ are shown by solid, dashed, and dotted lines, respectively.}
\end{center}
\end{figure}

The spectral energy distribution, $\nu F_\nu=x R(x)$, for the case $h_0$ is presented in Fig.~\ref{fig:TotalPowerSyn}. In contrary to the polarization, it decreases with the observation angle. This anticorrelation is seen from the comparison of the results  shown in Fig.~\ref{fig:SynPolarization} and \ref{fig:TotalPowerSynB}. Indeed, the broader distributions of magnetic field strength correspond to smaller polarization degree but larger intensity. The distribution $h_1$ gives a slower decrease of intensity than in the case $h_0$. In the case of $h_2$, as shown in Ref.~\cite{Kelner2013}, the spectral energy distribution falls down as a power-law distribution.

\begin{figure}
\begin{center}
\mbox{\includegraphics[width=0.5\textwidth,angle=0]{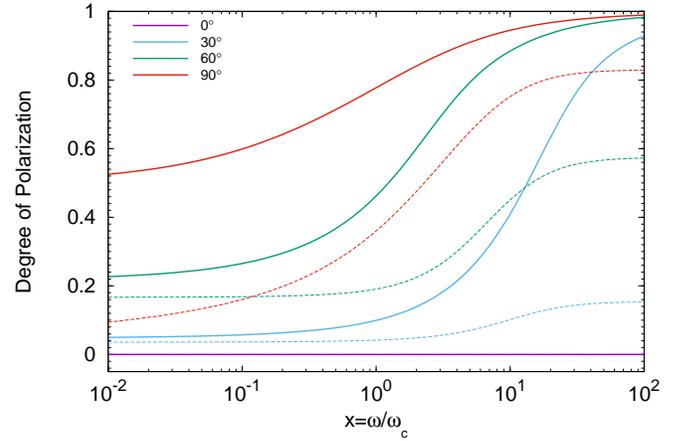}}
\caption{\label{fig:Comparison}Comparison of the degree of polarization for jitter (dashed lines) and synchrotron (solid lines) radiation for different observation
angles.}
\end{center}
\end{figure}

In Fig.~\ref{fig:Comparison} we compare the polarization degrees of the jitter and synchrotron radiation for different observation angles. At all angles and frequencies, the monoenergetic distribution of electrons produces more polarized radiation in the synchrotron than in the jitter regime. It can also be seen in Fig.~(\ref{fig:PolAngle}) where the low-frequency limit given by Eq.~(\ref{eq:jLowLimit}) is shown by dashed line. Note that this feature of  polarization of radiation by monoenergetic electrons, cannot be generalized for an arbitrary  distribution.
As it is shown in the next section the jitter radiation can be more polarized for certain electron distributions.

\subsection{Power-law spectra}
Eqs.~(\ref{eq:QRJit}) and (\ref{eq:pQR}) allow calculations of the radiation properties for arbitrary distributions of electrons. An important case is the power-law energy distribution of electrons. If one takes into account that the geometry of turbulent magnetic fields discussed in this paper can be realized in the shock waves where the particles are accelerated typically with a power-law distribution, the consideration of power-law  distribution of electrons is natural. In this section we derive the formulas for the polarization of jitter radiation for the power-law distributions of electrons and compare them with the expressions for the synchrotron radiation obtained in Ref.~\cite{Laing1980}.

For the power-law distribution over the Lorentz factor of electrons $\gamma$,
\begin{equation}
dN_e=C\gamma^{-p}d\gamma,
\end{equation}
the polarization tensor is 
\begin{equation}
(\bar{P}_{\b n \omega}(t))_{ik}=
\frac{e^4\langle \b B^2\rangle}{m^2c^4}\left(I_{j1}\delta_{ik}-I_{j2}(\delta_{ik}-2s'_is'_k)\right),
\end{equation}
where
\begin{eqnarray}
I_{j1,2}=\frac{C}{2}\frac{1}{(2\pi)^2}\left(\frac{\omega}{2c}\right)^{-\frac{p-1}{2}}\times\\\nonumber
\int\limits_{0}^{1}d\xi\, \xi^{\frac{p-3}{2}} F_{1,2}(\xi,\sigma)\int\limits_{0}^{\infty}
dq\, q^{\frac{p+1}{2}}\Psi(q).
\end{eqnarray}
It should be noted that in the case of a power-law electron energy distribution the dependence of the power spectrum of the jitter radiation on frequency is identical to the synchrotron spectrum, namely, $\sim \omega^{(p-1)/2}$. Introducing the spectral index of radiation $\zeta=(p-1)/2$, the degree of polarization of the jitter radiation is 
\begin{equation}\label{eq:PlawPol}
\Pi_{jp}=I_{j2}/I_{j1}=\int\limits_{0}^{1}d\xi\, \xi^{\zeta-1} F_2(\xi,\sigma)/\int\limits_{0}^{1}d\xi\, \xi^{\zeta-1} F_1(\xi,\sigma).
\end{equation}
The calculation of the integrals gives
\begin{equation}\label{eq:JitPowPol}
\Pi_{jp}=\frac{(1+\zeta)(2+\zeta)}{\zeta^2+3\zeta+4}
\left(1-\frac{2\sigma^2(1-\sigma^{1+\zeta})}{(1-\sigma^2)(\zeta+\sigma^{1+\zeta})}\right).
\end{equation}

Note this expression coincides with the formula given by Eq.~(\ref{eq:jHighLimit}) for the high-frequency limit
in the case of monoenergetic electron distribution. However, contrary to Eq.~(\ref{eq:jHighLimit}), this expression formally does not depend on the spectrum of the magnetic field turbulence. In the derivation of Eq.~\ref{eq:JitPowPol} we have neglected
the low-energy cut-off of the electron distribution $\gamma_0$, assuming $\gamma_0=0$. For the non-zero value of $\gamma_0$, in the limit $\omega_{pl}=\frac{3\omega}{4\omega_0}\delta_j\gg 1$, where $\omega_0$ is the characteristic frequency of the synchrotron radiation corresponding to the Lorentz factor $\gamma_0$, one can obtain the correction terms to the numerator and denominator in Eq.~(\ref{eq:PlawPol}). Specifically, in the case of the turbulent spectrum given by Eq.~(\ref{eq:JitTurbSpec}) we have
\begin{eqnarray*}\label{eq:asymptj}
I_{j1,2}\sim\int\limits_{0}^{1}d\xi\, \xi^{\zeta-1} F_{1,2}(\xi,\sigma)\int\limits_{0}^{\omega_{pl}/\lambda\xi}
dq\, q^{\zeta+1}\Psi(q)\underset{\omega_{pl}\rightarrow\infty}{\approx}\hspace{3.3cm}\\\nonumber
\approx\int\limits_{0}^{1}d\xi\, \xi^{\zeta-1} F_{1,2}(\xi,\sigma)\int\limits_{0}^{\infty}
dq\, q^{\zeta+1}\Psi(q)-\hspace{5.2cm}\\\nonumber
-\frac{A_{\alpha}\lambda^{1-\zeta}}{(\alpha-\zeta)\omega_{pl}^{\alpha-\zeta}}\int\limits_{0}^{1}d\xi\, \xi^{\alpha-1} F_{1,2}(\xi,\sigma).\hspace{3cm}\\\nonumber
\end{eqnarray*}
At large frequencies, the correction terms, which depend on the spectral index of the turbulence
$\alpha$, negligibly contribute to this expression, thus can be discarded.

The calculations of the degree of polarization of synchrotron radiation is \citep{Laing1980}
\begin{equation}\label{eq:SynPowPol}
\Pi_{sp}=I_{s2}/I_{s1},
\end{equation}
where
\begin{eqnarray}
I_{s1}=2\sigma^{\frac{\zeta+1}{2}}P_{\frac{\zeta+1}{2}}(t),\hspace{4cm}\\\nonumber
I_{s2}=\left(\frac{\zeta+1}{\zeta+5/3}\right)\sigma^{\frac{\zeta-1}{2}}\times\hspace{3.3cm}\\\nonumber
\times\left[\frac{2(1+\sigma^2)}{1+\zeta}P^1_{\frac{\zeta-1}{2}}(t)+(1-\sigma^2)P_{\frac{\zeta-1}{2}}(t)\right].
\end{eqnarray}
Here $t=(1+\sigma^2)/2\sigma$, and $P_q$ and $P^m_q$ are the Legendre and Associated Legendre functions, respectively.

\begin{figure}
\begin{center}
\mbox{\includegraphics[width=0.5\textwidth,angle=0]{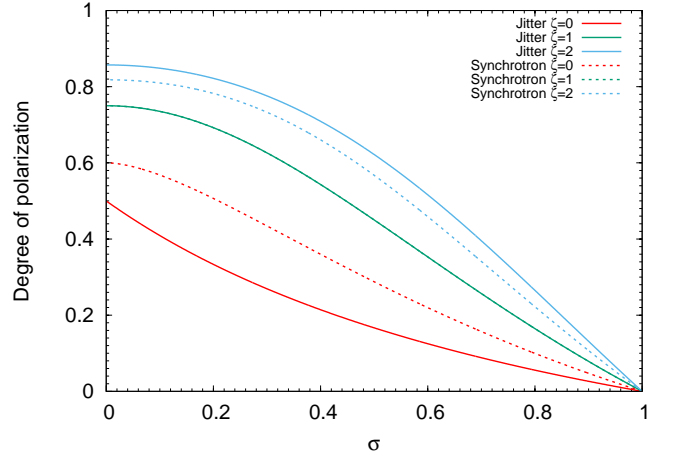}}
\caption{\label{fig:PowerPolar}The degree of polarization of the jitter (solid lines) and synchrotron radiation (dashed lines) as function of the cosine of the observation angle $\sigma=\b n\b s$. The results are shown for different values of index $\zeta$ of the power-law radiation spectrum produced by power-law energy distribution of
electrons. Note that at $\zeta=1$ the results  calculated for  two radiation regimes coincide, $\Pi=\frac{3}{4}\left( \frac{1-\sigma^2}{1+\sigma^2}\right)$.}
\end{center}
\end{figure}

The degrees of polarization as a function of the observation angle given by Eq.~(\ref{eq:JitPowPol}) for the jitter radiation (solid lines) and by Eq.~(\ref{eq:SynPowPol}) for the synchrotron
radiation (dashed lines) are presented in Fig.~\ref{fig:PowerPolar}. At spectral index of the radiation $\zeta=1$ both expressions for the polarization are equal to 
\begin{equation}
\Pi_{jp}(\zeta=1)=\Pi_{sp}(\zeta=1)=\frac{3}{4}\left(\frac{1-\sigma^2}{1+\sigma^2}\right).
\end{equation}
At $\zeta<1$ the polarization of the synchrotron radiation is higher. At $\zeta>1$ the polarization of the jitter radiation becomes larger at all observation angles.

\section{Conclusion}
In this paper, the polarization properties of the radiation produced by isotropically distributed electrons in a {\it turbulent} magnetic field with directions strictly parallel to the plane (so-called slab geometry) have been studied. We consider two extreme cases of the small and large turbulent scales. 

In the large-scale turbulent magnetic field ultrarelativistic electrons radiate in the regular synchrotron regime. The geometry of the field affects insignificantly the spectral energy distribution of radiation for a given turbulence spectrum. The intensity observed at different observation angles differs within a factor of two (see Fig.~\ref{fig:TotalPowerSyn}). The polarization is more sensitive to the observation angle. It changes from $0\%$, when the magnetic field plane is observed face-on to higher than $90\%$ when the plane is observed edge-on (Fig.~\ref{fig:SynPolarization}). Both the intensity and the polarization are sensitive to the distributions over the magnetic field strength (Figs.~\ref{fig:TotalPowerSynB} and \ref{fig:SynPolarization}). At high frequencies the radiation spectrum in the cutoff region falls down slower in the case of broader field distribution. At the same time the radiation becomes less polarized in turbulent field with a broader distribution.

In a small-scale turbulent magnetic field, the properties of radiation of electrons are substantially different from the properties of the synchrotron radiation. Namely, if the characteristic length of the turbulence $\lambda$ is 
smaller than the non-relativistic Larmor radius $R_L$, electrons emit in the jitter radiation regime. The jitter radiation is determined by the scale of turbulence and, therefore, by definition, occurs only in the turbulent media. In the slab geometry of turbulent magnetic field, which can be generated in the relativistic shock waves, e.g. by Weibel instability, we derived analytical formulae presented in Eqs.~(\ref{eq:JitTensor} and \ref{eq:QRJit}) in the tensor form. We derived also the spectral energy distribution of the jitter radiation field as a function of the observation angle $\theta$. After averaging over $\theta$, it naturally leads to the results derived in Ref.~\cite{Kelner2013} in the case of isotropic turbulence. 

The jitter radiation has distinct spectral features. The maximum of the spectral energy distribution is achieved at $\omega_{j}$ which is shifted $R_L/\lambda$ times towards higher frequencies compared to the position of the maximum of the synchrotron radiation. At high frequencies the spectrum has a power-law form; the slope depends on the spectrum of the magnetic turbulence (see Eq.~(\ref{eq:JHighPower})).  At low frequencies,  it is described by Eq.~(\ref{eq:assymp}) which tends to a constant.

As in the case of synchrotron radiation, the jitter radiation is not polarized when the magnetic field plane is observed face-on. It grows with increase of the angle from this direction. In general, the polarization of the jitter and synchrotron radiations have similar properties. In both regimes, it increases with the frequency and the observation angle. But they are different in details. In the case of the monoenergetic distribution of electrons, the polarization of the synchrotron radiation is higher than the polarization of the jitter radiation at all observation angles and frequencies. However, for the power-law distribution of elections the polarization of the jitter radiation can be higher.


\end{document}